\documentclass[final,5p,times,twocolumn]{elsarticle}
\usepackage{amssymb}
\usepackage[dvipsnames]{xcolor}
\usepackage{graphicx}% Include figure files
\usepackage{dcolumn}% Align table columns on decimal point
\usepackage{ulem}
\usepackage{bm}% bold math
\usepackage[colorlinks=true,linkcolor=blue,citecolor=blue]{hyperref}
\usepackage{siunitx}
\usepackage{xcolor}
\usepackage{gensymb}
\usepackage{parskip}
\usepackage{amsmath}
\usepackage{upgreek}
\usepackage{orcidlink}

\begin{document}
\setlength{\parindent}{2em}
\begin{frontmatter}

\title{Artificial ferroelectric-like hysteresis in antiferroelectrics with non-uniform disorder}


\author[inst1,equal]{Yi Zhang}
\author[inst1,equal]{Xinyu Zhang}
\author[inst2,equal]{Zihao Zheng}
\author[inst3,inst4]{Jiyang Xie}
\author[inst5,inst6]{Jing Lou}
\author[inst1]{Jiayi Qin}
\author[inst1]{Shanhu Wang}
\author[inst3,inst4]{Yang He}
\author[inst2]{Yifeng Du}
\author[inst2]{Bin Yang}
\author[inst1]{Xin Huang}
\author[inst1]{Huiping Han}
\author[inst1]{Yilin Wu}
\author[inst1]{Shuya Liu}
\author[inst1]{Afzal Khan}
\author[inst1]{Zhidong Li}
\author[inst1]{Qianxu Ye}
\author[inst1]{Sheng'an  Yang\orcidlink{0009-0009-1600-8852}}
\author[inst1,inst7]{Ji Ma\orcidlink{0000-0002-1308-3226}}
\author[inst1]{Hui Zhang}
\author[inst1]{Xiang Liu}
\author[inst1,inst7]{Qingming Chen}
\author[inst3,inst4]{Wanbiao Hu}
\author[inst8]{Jing Ma\orcidlink{0000-0003-0103-9858}}
\author[inst1,inst7]{Jianhong Yi}
\author[inst2,corresponding]{Jinming Guo\orcidlink{0000-0003-2556-709X}}
\author[inst5,inst6,inst9]{Shou Peng}
\author[inst9,corresponding]{Hao Pan\orcidlink{0000-0001-9620-3440}}
\author[inst1,corresponding]{Liang Wu\orcidlink{0000-0003-1030-6997}}
\author[inst8]{Ce-Wen Nan\orcidlink{0000-0002-3261-4053}}

\affiliation[inst1]{organization={Faculty of Materials Science and Engineering, Kunming University of Science and Technology},%Department and Organization
            % addressline={Address One}, 
            city={Kunming},
            postcode={650093}, 
            state={Yunnan},
            country={China}}
\affiliation[inst2]{organization={Ministry-of-Education Key Laboratory of Green Preparation and Application for Functional Materials, School of Materials Science and Engineering, Hubei University},%Department and Organization
            % addressline={Address One}, 
            city={Wuhan},
            postcode={430062}, 
            state={Hubei},
            country={China}}            
\affiliation[inst3]{organization={Yunnan Key Laboratory of Electromagnetic Materials and Devices, National Center for International Research on Photoelectric and Energy Materials, School of Materials and Energy, Yunnan University},%Department and Organization
            % addressline={Address One}, 
            city={Kunming},
            postcode={650091}, 
            state={Yunnan},
            country={China}}  
\affiliation[inst4]{organization={Electron Microscopy Center, Yunnan University},%Department and Organization
            % addressline={Address One}, 
            city={Kunming},
            postcode={650091}, 
            state={Yunnan},
            country={China}}  
\affiliation[inst5]{organization={Shenzhen Triumph Science and Technology Engineering Co. Ltd.},%Department and Organization
            % addressline={Address One}, 
            city={Shenzhen},
            postcode={518054}, 
            state={Guangdong},
            country={China}}         
\affiliation[inst6]{organization={State Key Laboratory of Advanced Glass Materials, CNBM Research Institute for Advanced Glass Materials Group Co., Ltd.},%Department and Organization
            % addressline={Address One}, 
            city={Bengbu},
            postcode={233000}, 
            state={Anhui},
            country={China}}    
\affiliation[inst7]{organization={Southwest United Graduate School},%Department and Organization
            % addressline={Address One}, 
            city={Kunming},
            postcode={650093}, 
            state={Yunnan},
            country={China}}  
\affiliation[inst8]{organization={School of Materials Science and Engineering, Tsinghua University},%Department and Organization
            % addressline={Address One}, 
            city={Beijing},
            postcode={100084}, 
            % state={Yunnan},
            country={China}}              
\affiliation[inst9]{organization={School of Advanced Materials, Shenzhen Graduate School, Peking University},%Department and Organization
            % addressline={Address One}, 
            city={Shenzhen},
            postcode={518055}, 
            state={Guangdong},
            country={China}}

\affiliation[equal]{country={These authors contributed equally}%Department and Organization
            % addressline={Address Two}, 
            % city={City Two},
            % postcode={22222}, 
            % state={State Two},
            % country={Country Two}
            }

\affiliation[corresponding]{country={guojinming@hubu.edu.cn; panh@pku.edu.cn; liangwu@kust.edu.cn}}

\begin{abstract}
Antiferroelectrics exhibit unique double-hysteresis polarization loops, which have garnered significant attention due to their potential applications such as energy storage, electromechanical transduction, as well as synapse devices.
However, numerous antiferroelectric materials have been reported to display signs of hysteresis loops resembling those of ferroelectric materials, and a comprehensive understanding remains elusive. 
In this work, we provide a phenomenological model that reproduces such widely observed artificial ferroelectric hysteresis with a superposition of numerous disordered antiferroelectric loops that have varying antiferroelectric-to-ferroelectric transition fields, particularly when these field ranges intersect. Experimentally, we realized such artificial ferroelectric-like hysteresis loops in the prototypical antiferroelectric PbZrO$_3$ and PbHfO$_3$ thin films, by introducing non-uniform local disorder (e.g., defects) via fine-tuning of the film growth conditions. These ferroelectric-like states are capable of persisting for several hours prior to transitioning back into the thermodynamically stable antiferroelectric ground state. Those results provide insights into the fundamental impact of disorder on the AFE properties and new possibilities of disorder-tailored functions. 
\end{abstract}


\end{frontmatter}

\renewcommand{\figurename}{}
\renewcommand{\thefigure}{Fig. \arabic{figure}}

\section{Introduction}
\label{sec:sample1}
Local disorder, such as defects in otherwise perfect crystals or materials, has long been acknowledged as a formidable foe for theoreticians, while at the same time it can serve as a crucial (but often undervalued) knob for experimentalists to achieve enhanced or new functionalities. Celebrated examples include the donor and/or acceptor impurities in semiconductors, defect pinning enhanced critical current density in superconductors {\cite{Larbalestier2001, Haugan2004}}, and defect-induced luminescence in LEDs {\cite{Nakamura1994, Nakamura2013}}. Such local disorder has also been demonstrated to make noticeable impacts on the behaviors of ferroelectric (FE) materials, which have spontaneous and voltage-switchable electric dipoles (with a hysteresis polarization-electric field ($P$-$E$) loop) and have been extensively studied as promising candidates for nonvolatile electronics, piezoelectrics, high-permittivity dielectrics, etc {\cite{Martin2016, Feng2020}}. The local disorder (defects) can directly interact with the FE dipoles and alter the polarization switching {\cite{Fernandez2022}}, causing performance degradation such as fatigue, aging, and imprint {\cite{Baek2011, Damodaran2014}}. On the other hand, works (such as defect engineering) have also shown that gaining control of such disorder can provide a new degree of freedom to realize exotic polarization behaviors and even improved properties. For instance, antisite point defects were found to induce FE features from nonpolar materials {\cite{Lee2015, Ning2021}}, while the FE polarization retention could be further enhanced by defect pinning on FE domain walls {\cite{Zhang2020}}.  The microscopic FE domain structures have also been demonstrated to be finely engineered via structural defect ordering/disordering {\cite{Gradauskaite2022}}, defect-related electric boundary conditions {\cite{Liu2022}}, as well as built-in fields induced by point defects {\cite{Weymann2020, Sarott2023}}.

In comparison to the FE materials, the role of disorder in their antiferroelectric (AFE) counterparts has not been given as much consideration. AFE materials, such as the prototypical PbZrO$_3$ (PZO) and PbHfO$_3$ (PHO), have an antipolar ground state with antiparallel local dipoles that can be switched to parallel (i.e., the polar FE state) by external electric fields, which gives rise to a double-hysteresis $P$-$E$ loop \cite{Si2024}. This characteristic renders AFE highly promising for applications in capacitive energy storage {\cite{Acharya2022, Luo2023, Zhang2024-2, Xu2025}}, electromechanical transduction {\cite{Pan2024}}, synaptic devices {\cite{Cao2022}}, and thermal switch \cite{Liu2023-2}. Recently, studies have demonstrated that, similar to the case of FEs, various kinds of local disorder such as point defects {\cite{Wei2021}}, dislocations {\cite{Chaudhuri2011}}, and antiphase boundaries {\cite{Liu2023}} can be introduced into AFEs. As such, the energy competition between the AFE and FE states can be delicately altered, causing a phase transition to a new ground FE state with the evolution of the polarization behavior from double-hysteresis to single-hysteresis loops {\cite{Gao2017}}. More recently, it was revealed that, even when the ground state remained to be AFE, single-hysteresis polarization loops could also be observed with the point defects pinning the AFE-to-FE phase boundary and delaying the phase transition kinetics {\cite{Pan2023}}. Here we define such a behavior as artificial FE-like, differentiating it from real FE materials that possess a polar FE ground state. Despite these efforts and advances in AFEs, the fundamental understanding from the perspective of disorder has yet to be formulated to elucidate the evolution of AFE polarization behaviors, which is desirable for the tailoring of AFE materials for improved properties and practical applications.

In this work, we present a phenomenological model to mimic the artificial FE-like hysteresis observed in the quintessential AFE materials (PZO and PHO) with non-uniform disorder. The increase of disorder level is found to smear the AFE double-hysteresis $P$-$E$ loops, and further superposition of loops with varied disorder levels (mimicking the real material with non-uniform disorder), can generate an artificial FE-like single-hysteresis $P$-$E$ loop. Such an evolution is experimentally observed in the PZO and PHO thin films that are coherently grown by pulsed laser deposition (PLD) on La-doped BaSnO$_3$ (LBSO, $a_\text{pc} = 4.116$ {\AA}) bottom electrodes (which also served as a strain-relaxed buffer layer on SrTiO$_3$(001) (STO, $a_\text{pc} = 3.905$ {\AA}) substrates) {\cite{Eom2022}}. The disorder level in the films, e.g., Pb volatility and the crystallinity, is systematically controlled by the growth temperature and growth oxygen pressure, which causes gradual smearing of the AFE double-hysteresis polarization behavior and eventually an artificial FE-like polarization switching.  Such an artificial FE phase is further found to be metastable, which can retain for hours but eventually revert back to the AFE ground state at room temperature. These observations and understanding highlight the fundamental impact of disorder on the AFE properties and point out the new possibilities of disorder-tailored functions.   

% \appendix

\section{Experimental} \label{sec3}

\noindent
\textit{2.1 Simulations of polarization hysteresis dependent on disorder intensity}:
Simulations of the polarization hysteresis were performed based on the thermodynamic energy landscapes that were disturbed by various levels of non-uniform disorder ``noise", which used a Python script specifically written for this study. The detailed parameters of the energy landscapes and the disorder can be found in the next section.

\noindent
\textit{2.2 Fabrication of epitaxial films}:
Epitaxial AFE PZO and PHO thin films were fabricated using pulsed laser deposition (PLD, Anhui Epitaxy Technology) with a KrF excimer laser (with a wavelength of 248 nm).  Single-phase ceramic targets were employed for the deposition, with the ceramic synthesis detailed in Ref.\cite{Zhang2024}. The LBSO bottom electrode was deposited at a growth temperature of 850 $^{\circ}$C (measured using a pyrometer), with an oxygen background pressure of 100 mTorr, a laser energy density of 1.0 J/cm$^{2}$ (with a repetition rate of 3 Hz),  and a target-substrate distance of 5.5 cm.  The PZO layer ($\sim$ 250 nm) was subsequently grown at various temperatures (500, 520, 540, 560, 580, and 600 $^{\circ}$C) with an oxygen pressure of 26.6 Pa, a laser fluence of 1.3 J/cm$^{2}$, and a laser repetition rate of 5 Hz.
The PHO layer ($\sim$ 400 nm) was grown at various temperatures (from 540 to 660 $^{\circ}$C, with an increment of 20 $^{\circ}$C) and with varied oxygen pressure (from 20 to 120 Pa, with an increment of 20 Pa), using a laser fluence of 0.7 J/cm$^{2}$ and a laser repetition rate of 5 Hz. PHO films were also grown at a temperature of 500 $^{\circ}$C and an oxygen pressure of 8 Pa (conditions considerably deviating from the above parameters) to deliberately incorporate a high degree of disorder. The above as-grown heterostructures were then cooled to room temperature at a rate of 15 $^{\circ}$C min$^{-1}$ in an oxygen pressure of 700 Torr. Finally,  circular Au top electrodes with a diameter of 100 $\upmu$m were deposited through a shadow mask by an ion sputter.

\noindent
\textit{2.3 X-ray diffraction}:
The structural properties were characterized using a high-resolution X-ray diffractometer (XRD, Rigaku SmartLab) with the X-ray from Cu $K_\alpha$ radiation ($\lambda =1.540598$ Å) and a two-bounce Ge(220) monochromator. 

\noindent
\textit{2.4 Polarization--electric-field hysteresis loops}:
Polarization-electric field hysteresis loops were measured with a bipolar triangular voltage profile at a frequency of 1 kHz using a ferroelectric tester (TF2000, aixACCT). 

\noindent
\textit{2.5 Piezoresponse force microscopy}:
Piezoresponse force microscopy (PFM) was performed using a commercially available scanning probe microscope (MFP-3D, Asylum Research--Oxford Instruments). To obtain the PFM images, dual amplitude resonance tracking (DART) mode was employed to maintain the operational frequency near the contact resonance frequency. A conductive Pt/Ir-coated probe tip (NanoSensor PPP-EFM) was used, featuring a spring constant of $k=2.8$ N/m and a free air resonance frequency of $f=75$ kHz. In all PFM characterizations, the bias was applied at the tip. Contact was made to the bottom LBSO electrode for grounding in the PFM studies.

\noindent
\textit{2.6 Scanning transmission electron microscopy}:
Scanning transmission electron microscopy (STEM) specimens were prepared by a dual-beam Focused Ion Beam (FIB, Helios 5 UC, ThermoFisher Scientific). The target cutting areas in thin films were first deposited with tungsten protection layers with a thickness of 2 $\upmu$m, and then the surrounding regions were removed by accelerated Ga ions. The specimens were lifted out by a tungsten micromanipulator and then bonded on commercial Cu grids. The accelerating voltage and beam current of Ga ions were 30 kV and 2.5 nA for initial lamella cutting, and both values were decreased gradually for further precise thinning down; finally, 2 kV and 40 pA were used for the removal of the surface amorphous layer. The final thicknesses of TEM lamellas were about 30--40 nm. The atomic-scale structures and chemical compositions of TEM lamellas were investigated by a double spherical aberration (Cs) corrected scanning transmission electron microscopy (STEM, Spectra 300, ThermoFisher Scientific) equipped with probe and image correctors and operated at 300 kV. The probe convergence angle and high-angle annular dark-field (HAADF) collection angles were 25 mrad and 39--200 mrad, respectively. The polarization vector and GPA data were obtained by calculating the displacement of B-site atoms from the center of the four neighboring A-site atoms, and the rotation angle via the customized Matlab scripts.

\section{Results and Discussion}

{\color{black}We first used a phenomenological model of AFE {\cite{Kittel1951, Cross1967, Okada1969, Hoffmann2022}} to show how disorder can affect the free energy density ($f$)-polarization ($P$) landscape and thus the $P$-$E$ switching behaviors. 

\begin{figure}[h!]
\centering
\includegraphics[width=\linewidth]{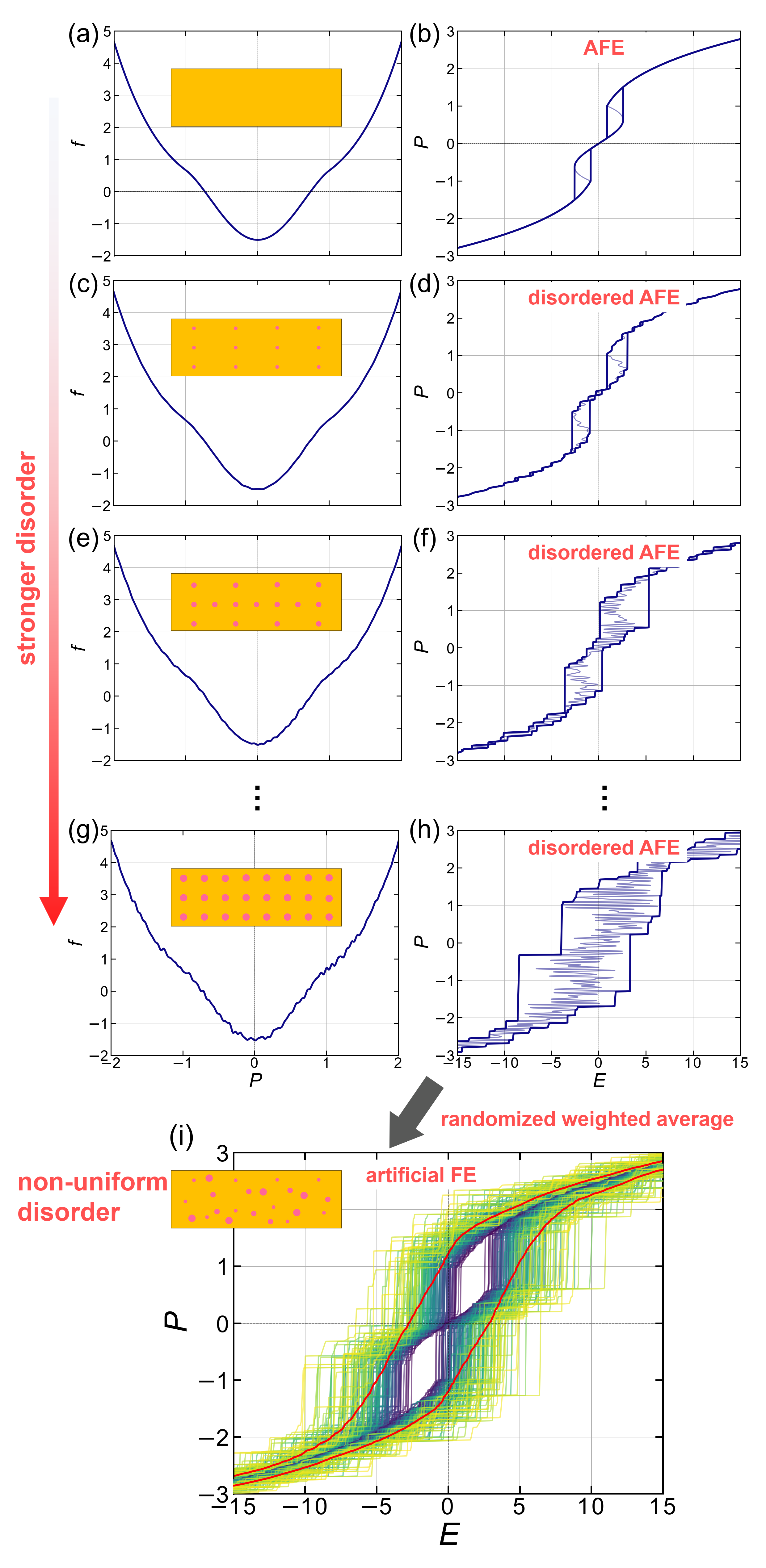}
\caption{Landau free energy density $f$ vs. polarization $P$ in disorder-free materials (a) and its corresponding $P$ vs. electric field $E$ ($P$-$E$) loop with an AFE double-hysteresis feature (b). (c, e, g) The $f$-$P$ curves of disordered materials, with increase of the disorder strength, and their corresponding $P$-$E$ loops (d, f, h). The insets in (a, c, e, g) depict the schematic representation of the disorder density and strength. (i) The overall weighted $P$-$E$ loop in the presence of non-uniform disorder tends to blur the sharp transition characteristic, resulting in a $P$-$E$ loop with single-hysteresis FE-like feature.}
\label{Fig1}
\end{figure}

{\color{black}
The $f$ of the AFE phase can be described as
\begin{equation}
f = 
\begin{cases}
(\alpha - 3\beta)P^2 - 8\zeta P^4 - \frac{\beta^2}{4\zeta} - t \times \text{noise}(P), &  \\ \hspace{40mm} \text{when} \ |P| < P_n \ (\text{non-polar}) \\
\alpha P^2 + \zeta P^4 - t \times \text{noise}(P), &  \\ \hspace{40mm}\text{when} \
|P| \geq P_n \ (\text{polar})
\end{cases}
\end{equation}
where $\alpha$, $\beta$, and $\zeta$ represent the Landau coefficients of the symmetry-permitted terms, $P=P_a + P_b$ is macroscopically measurable polarization (one of the AFE order
parameters, the other is the staggered polarization $Q= P_a - P_b$, and the $P_a$ and $P_b$ are the antiparallel sublattice polarizations in AFEs), $P_n = \sqrt{-\beta/(6\zeta)}$ specifies the $P$-boundary separating the non-polar reference phase from the polar phase. The detailed derivation process can be found in Ref.\cite{Hoffmann2022}.}
Additionally, we incorporated a noise energy term ($P$-dependent random number in the range of $[-1, 1]$) to model the fluctuations in free energy density caused by disorders (such as defect pinning), with its magnitude controlled by $t$.
In this scenario, the $P$-$E$ loop is obtained by calculating $E$ as the first derivative of $f$ with respect to $P$,
\begin{equation}
E = 
\begin{cases} 
2(\alpha - 3\beta) P - 32 \zeta P^3 - t \frac{\partial \text{noise}(P)}{\partial P}, & \text{when } |P| < P_n \text{ (non-polar)} \\
2\alpha P + 4 \zeta P^3 -t \frac{\partial \text{noise}(P)}{\partial P}, & \text{when } |P| \geq P_n \text{ (polar)}
\end{cases}
\end{equation}

Without loss of generality, we take $\alpha$, $\beta$, and $\zeta$ to be 0.1, $-1$, and $\frac{1}{6}$, respectively.  

For an ideal material (without any disorder, i.e., $t=0$), an AFE-type free energy landscape and $P$-$E$ loop are seen in \ref{Fig1}a and \ref{Fig1}b, respectively. Note that the forbidden region of negative free energy curvature with $\partial ^2 f/ \partial P^2 < 0$ is thermodynamically unstable; the $P$-$E$ loop is thus featured with a double-hysteresis loop.

\begin{figure*}[t]
\includegraphics[width=\linewidth]{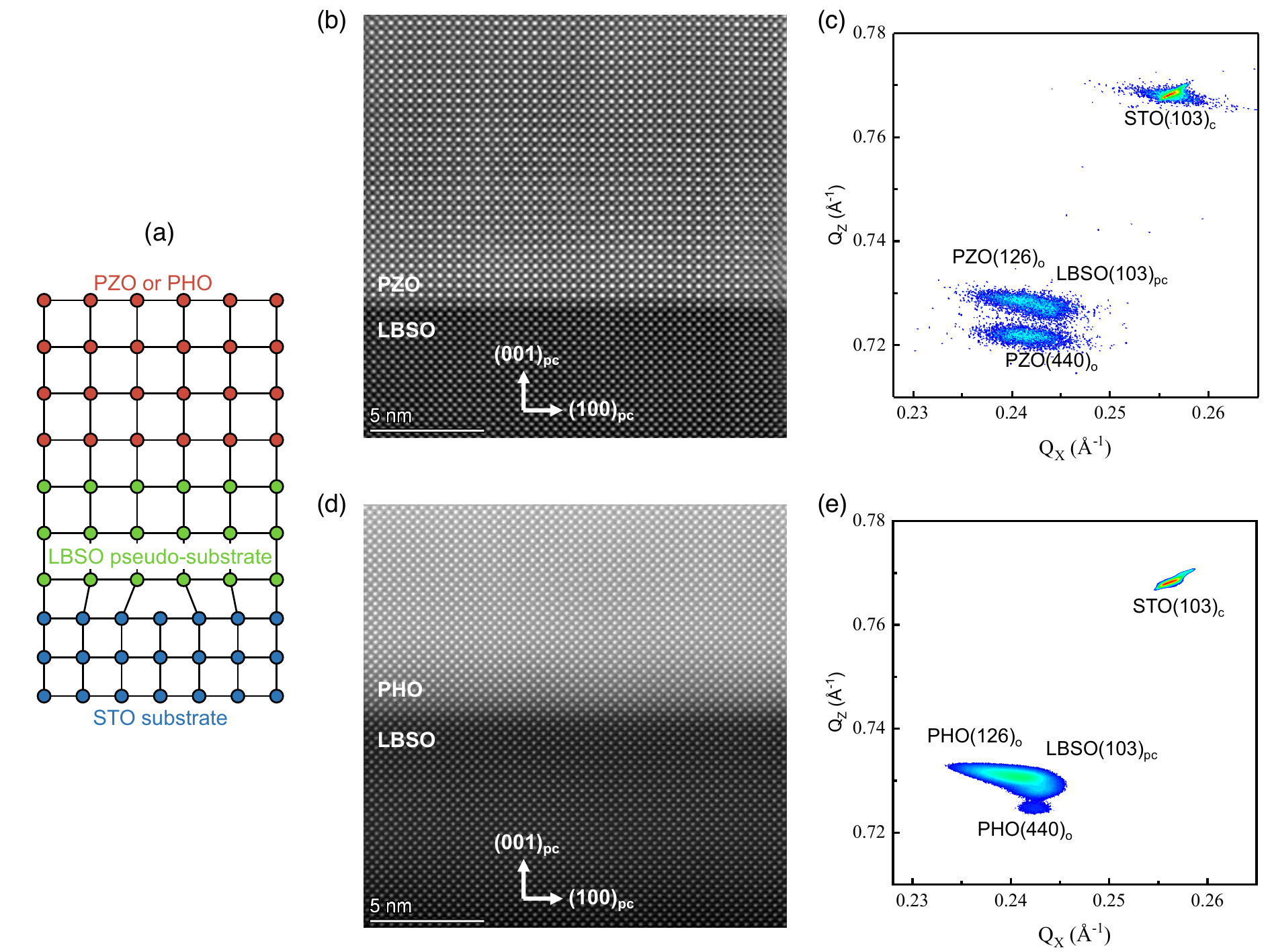}
\caption{(a) Schematic figure of coherent epitaxy of PZO or PHO on a LBSO pseudo-substrate. (b) The HAADF image of dislocation-free PZO/LBSO interface. (c) The RSM of PZO/LBSO/STO(001) heterostructure reveals the expected epitaxial relationship as schematically shown in (a).}
\label{Fig2}
\end{figure*}

When disorder intensifies (with $t$ spanning from 0 (\ref{Fig1}a) to 0.03 (\ref{Fig1}g), the $f$-$P$ energy curves accordingly become more noisy. Metastable local minimal states emerge, representing the pinning effect of disorders (such as point defects, random fields, dislocations, etc.) on the polarization (i.e., pinning of domain walls at the microscopic scale), which causes intermediate states in the $P$-$E$ curves. The increased strength of disorder (schematically illustrated by the density and size of the red dots in \ref{Fig1}c,e,g) enhances the depth of the local minima in the $f$-$P$ curves. Accordingly, the $P$-$E$ curve becomes so distorted that the double-hysteresis feature almost vanishes (\ref{Fig1}f,h).

One step further, considering that in real materials, the distribution of disorder should be inevitably non-uniform to a certain extent (as schematically shown in \ref{Fig1}i), the overall $P$-$E$ loop (depicted as the red curve in \ref{Fig1}i), representing a weighted average of all $P$-$E$ loops with varied strength of disorder, reveals a smearing of the distinct polarization switching steps, resulting in a continuous single-hysteresis switching.
Such a gradual increase in disorder strength along with a widening of its distribution, which diminishes clarity in the AFE double-hysteresis feature and renders the $P$-$E$ loop increasingly FE-like (with single hysteresis), nicely replicates the trend observed in previous study on AFE materials {\cite{Pan2023}}, highlighting how the nonuniform disorder (noise) can distort the energy landscape and the hysteretic polarization behaviors, i.e., from a typical AFE double-hysteresis loop to an FE-like single-hysteresis loop, without the stabilization of a ground-state FE phase.}
{\color{black} Actually, we can simply use the FE order parameter to pictorially approximate the AFE-like double-hysteresis loop, and represent the energy noise caused by disorder with trigonometric functions (the noise term can be represented as various combinations of trigonometric series via Fourier transformation). This schematically illustrates the general trend of transitioning from AFE to pinched $P$-$E$ and subsequently to FE-like behavior, as depicted in Figure S1, Supplementary Materials.}

\begin{figure*}[t]
\includegraphics[width=\linewidth]{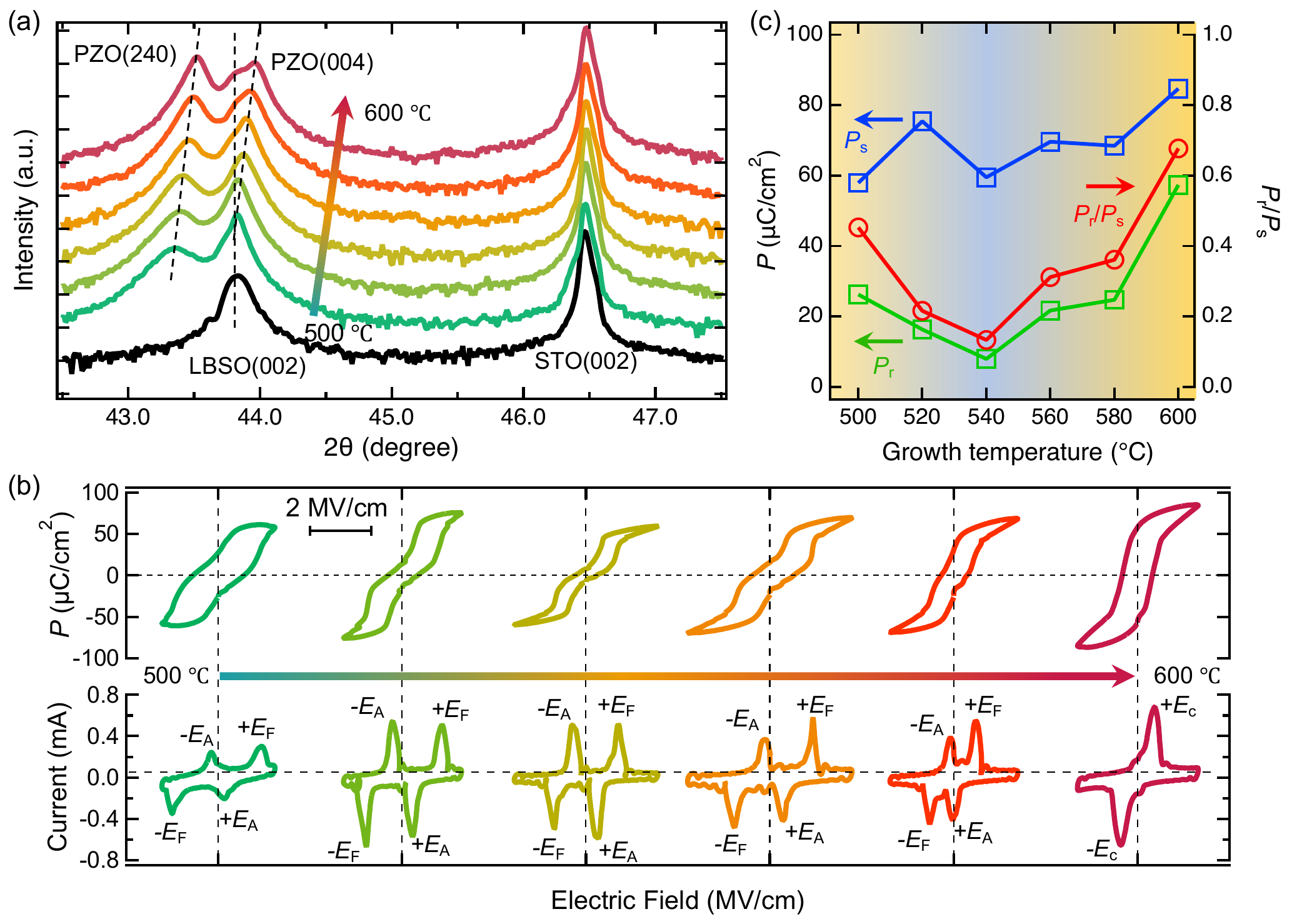}
\caption{(a) $2\theta$--$\omega$ scans for PZO/LBSO/STO(001)$_\text{c}$ grown at 500--600 $^{\circ}$C, exhibiting a gradual shift of the PZO peak towards higher $2\theta$ angle with increasing growth temperature, as well as a more PZO(240)$_\text{o}$ orientation. The black line was obtained from a bare LBSO thin film, and a vertical offset was added to the XRD data for clarity. (b) the $P$-$E$ loops and the switching current curves (differential of polarization with field) of the same samples in (a). Horizontal offsets were added for clarity. The well-defined AFE PZO was developed at a temperature of 540 $^{\circ}$C, which can be readily converted into a artificial FE by altering the growth temperatures. By meticulously examining the switching current, even in the most FE-like $P$-$E$ loop, two minor peaks near $\pm E_\text{A}$ are still observable, suggesting an AFE nature. (c) The remnant polarization, $P_\text{r}$ (at zero electric field), saturation polarization, $P_\text{s}$ (at maximized electric field) and their ratio $P_\text{r}/P_\text{s}$ at varying growth conditions. The smaller ratio indicates more ``pure” AFE.}
\label{PZO}
\end{figure*}

To experimentally verify this phenomenological model and investigate how disorder (such as defects) affects the polarization properties, a collection of PZO/LBSO/STO(001)$_\text{c}$ and PHO/LBSO/STO(001)$_\text{c}$ thin films were fabricated by PLD with deliberately varied parameters (including growth temperature and oxygen pressure; details see Experimental section). The premise to elaborate on the above issue is to synthesize AFE thin films with controllable degrees of disorder on the basis of the least disordered one. 
{\color{black}For the prototypical isostructural AFE PZO and PHO model systems, which possess the same space group $Pbam$ with lattice parameters of $a_\text{o}(\text{PZO}) \sim 5.872$ {\AA}, $b_\text{o}(\text{PZO}) \sim 11.744$ {\AA}, $c_\text{o}(\text{PZO}) \sim 8.202$ {\AA}; and $a_\text{o}(\text{PHO}) \sim 5.854$ {\AA}, $b_\text{o}(\text{PHO}) \sim 11.694$ {\AA}, $c_\text{o}(\text{PHO}) \sim 8.191$ {\AA}) {\cite{Acharya2022, Liu2023-2}}, respectively.} Their lattice parameters based on pseudocubic unit cells can be roughly calculated as 
$ \sim  \sqrt[3]{a_\text{o}(\text{PZO})b_\text{o}(\text{PZO})c_\text{o}(\text{PZO})/8} \sim 4.135$ {\AA} and 
$ \sim  \sqrt[3]{a_\text{o}(\text{PHO})b_\text{o}(\text{PHO})c_\text{o}(\text{PHO})/8} \sim 4.123$ {\AA}, respectively. Such lattice parameters are remarkably larger than those ($\sim$ 3.7--4.0 {\AA}) of typical commercially available perovskite substrates {\cite{Uecker2016}}.
Consequently, PZO and PHO have yet to be synthesized in a coherent epitaxial form, which would inherently bear uncontrollable disorder of dislocations that can suppress the polarization {\cite{Alpay2004}}. Therefore, dislocation-free PZO and PHO films are of vital importance to study the effect of disorder, and to optimize their properties for potential applications. As such, an LBSO layer with a lattice constant of $4.116$ {\AA} was used as bottom electrodes. As schematically illustrated in \ref{Fig2}a, the significant lattice mismatch ($>5\%$) between LBSO and the STO substrate causes interfacial dislocations, which remain only in the LBSO layer (Figure S2, Supplementary Materials). In contrast, the lattice mismatch between PZO (or PHO) and LBSO is much smaller ($<1\%$), which facilitates a coherent growth of PZO (or PHO) on LBSO without the formation of dislocations. This is confirmed by structural characterizations with HAADF--STEM and RSM. As shown in representative STEM images of the PZO/LBSO/STO and PZO/LBSO/STO heterostructures (\ref{Fig2}b,d), the interface between PZO (or PHO) and LBSO is well-defined and free from dislocations. The RSM result around the STO(103)$_\text{c}$ diffraction condition further reveals the coherent epitaxy of PZO (or PHO) on LBSO, indicated by the near-equivalent in-plane lattice constants of PZO(126)$_\text{o}$, PZO(440)$_\text{o}$, and LBSO(103)$_\text{pc}$ diffraction peaks, which all relax from the STO substrate (The subscripts ``c", ``o'', and ``pc" denote cubic, orthorhombic, and pseudocubic, respectively; \ref{Fig2}c,e). 

Based on such dislocation-free PZO and PHO heterostructures, we first studied how the growth temperature introduces disorder into the AFE films and impacts their polarization behaviors. The structural evolution of the PZO/LBSO/STO(001)$_\text{c}$ heterostructures with growth temperature increasing from 500 to 600  $^{\circ}$C  (20 $^{\circ}$C per step) was checked by $2\theta$--$\omega$ XRD scans (Experimental), as seen in \ref{PZO}a (a wider range of $2\theta$--$\omega$ scans can be found in Figure S3, Supplementary Materials). All the PZO thin films exhibit a mixture of  (240)$_\text{o}$ and (004)$_\text{o}$ orientations, in accordance with previous reports {\cite{Pan2023, Dufour2023}}.
Note that the strain values for the (240)$_\text{o}$ and (004)$_\text{o}$ orientations of PZO on the LBSO layer are
\begin{align}
    f_{(240)_\text{o}} &= \frac{\frac{\sqrt{\left(\frac{a_\text{o}(\text{PZO})}{\sqrt{2}}\right)^2 + \left(\frac{c_\text{o}(\text{PZO})}{2}\right)^2}}{\sqrt{2}}-a_\text{pc}(\text{LBSO})}{\frac{\sqrt{\left(\frac{a_\text{o}(\text{PZO})}{\sqrt{2}}\right)^2 + \left(\frac{c_\text{o}(\text{PZO})}{2}\right)^2}}{\sqrt{2}}} \approx 0.27\%, \\
    f_{(004)_\text{o}} &= \frac{\frac{a_\text{o}(\text{PZO})}{\sqrt{2}}-a_\text{pc}(\text{LBSO})}{\frac{a_\text{o}(\text{PZO})}{\sqrt{2}}}  \approx 0.87\%,
\end{align}
respectively (given that $a_\text{pc}(\text{PZO}) \approx b_\text{pc}(\text{PZO}) \approx 4.152$ \AA, $c_\text{pc}(\text{PZO}) \approx 4.101$ \AA). The small lattice mismatch leads to the coexistence of the two orientations, which are further influenced by the growth conditions. With the increased growth temperature, while the in-plane lattice of PZO should be constrained by the LBSO layer, the out-of-plane lattice parameter of PZO decreases progressively, evidenced by the PZO(240)$_\text{o}$ and PZO (004)$_\text{o}$ peaks both moving towards higher XRD angles. 
{\color{black}This observation can be attributed to Pb concentration gradient with excess Pb cations accumulating near the film surface at higher growth temperatures, which is consistent with earlier findings of growth-temperature-driven Pb off-stoichiometry in PbTiO$_3$ {\cite{Weymann2020}} and Pb(Zr$_{0.2}$Ti$_{0.8}$)O$_3$ films {\cite{Sarott2023}}.}
As the lattices shrink with increasing growth temperature, the PZO(240)$_\text{o}$ orientation becomes more coherent (with minimal lattice mismatch) with the substrate, therefore shows continuously increasing peak intensity compared to that of  PZO(004)$_\text{o}$, as shown in \ref{PZO}a.

As highlighted in previous reports {\cite{Weymann2020, Sarott2023}}, in lead-based films with increasing growth temperature, the enhanced volatility and mobility of Pb cause an excess of Pb near the film surface, thus a gradient of Pb concentration across the film thickness. Such an imperfection (from point defects) can act as a key source of the continuous change of the disorder level in our phenomenological model. As such, $P$-$E$ loops of the heterostructures grown at varied temperatures were measured (\ref{PZO}b). As immediately seen, the PZO heterostructure grown at 540 $^{\circ}$C shows the most typical AFE polarization switching with a double-hysteresis loop. With either increasing or decreasing growth temperatures, a shift towards a pinched polarization feature, and eventually to an artificial FE-like $P$-$E$ loop (especially at higher temperatures) that aligns well with our theoretical model (\ref{Fig1} and S1, Supplementary Materials), was observed. 
{\color{black}It is noted that the polarization value for the FE-like $P$-$E$ loops observed in our study is higher than the values of our optimal AFE PZO films (grown at 540 $^{\circ}$C) and those reported in the literature \cite{Liu2023-2, Gao2017, Dufour2023, Lisenkov2020}. This occurs because leakage currents rise when growth conditions differ from optimal, which does not truly enhance polarization but instead results in  overestimation in polarization measurements. This contribution is evidenced by the large ``non-overlapping" in the $P$-$E$ loop observed in the high-field range of the increasing and decreasing electric field. Even as the electric field is reduced, the polarization continues to increase.}
The switching current curves, which were derived as the differential of polarization with field, also demonstrate such an evolution from AFE switching (with the AFE-to-FE transition peaks $\pm E_\text{F}$ and FE-to-AFE transition peaks $\pm E_\text{A}$) to FE-like switching (with only two major switching coercive peaks $\pm E_\text{c}$). This evolution is also reflected by the evolution of the $P_\text{r}/P_\text{s}$ ratio (red curve in \ref{PZO}c), which is the lowest for the growth temperature of 540 $^{\circ}$C. Deviating from this optimal growth condition can lead to an increase in  $P_\text{r}/P_\text{s}$  with enhanced defect complexity (such as defect concentration, category, distribution, etc.) in the thin films that can pin the AFE-FE phase boundary and delay the kinetics of backswitching from FE to AFE phase, causing an increase of $P_\text{r}$ {\cite{Pan2023}}. The intrinsic inhomogeneity of these defects, such as Pb off-stoichiometry, oxygen vacancies, and the associated defect complexes/clusters, broadens the distribution of disorder strength $t$, which is equivalent to broadening the distribution of pinning energy. This accounts for the observed evolution of $P$-$E$ loops from typical AFE to FE-like behavior within the framework of our theoretical model.

\begin{figure*}[h!]
\centering
\includegraphics[width=\linewidth]{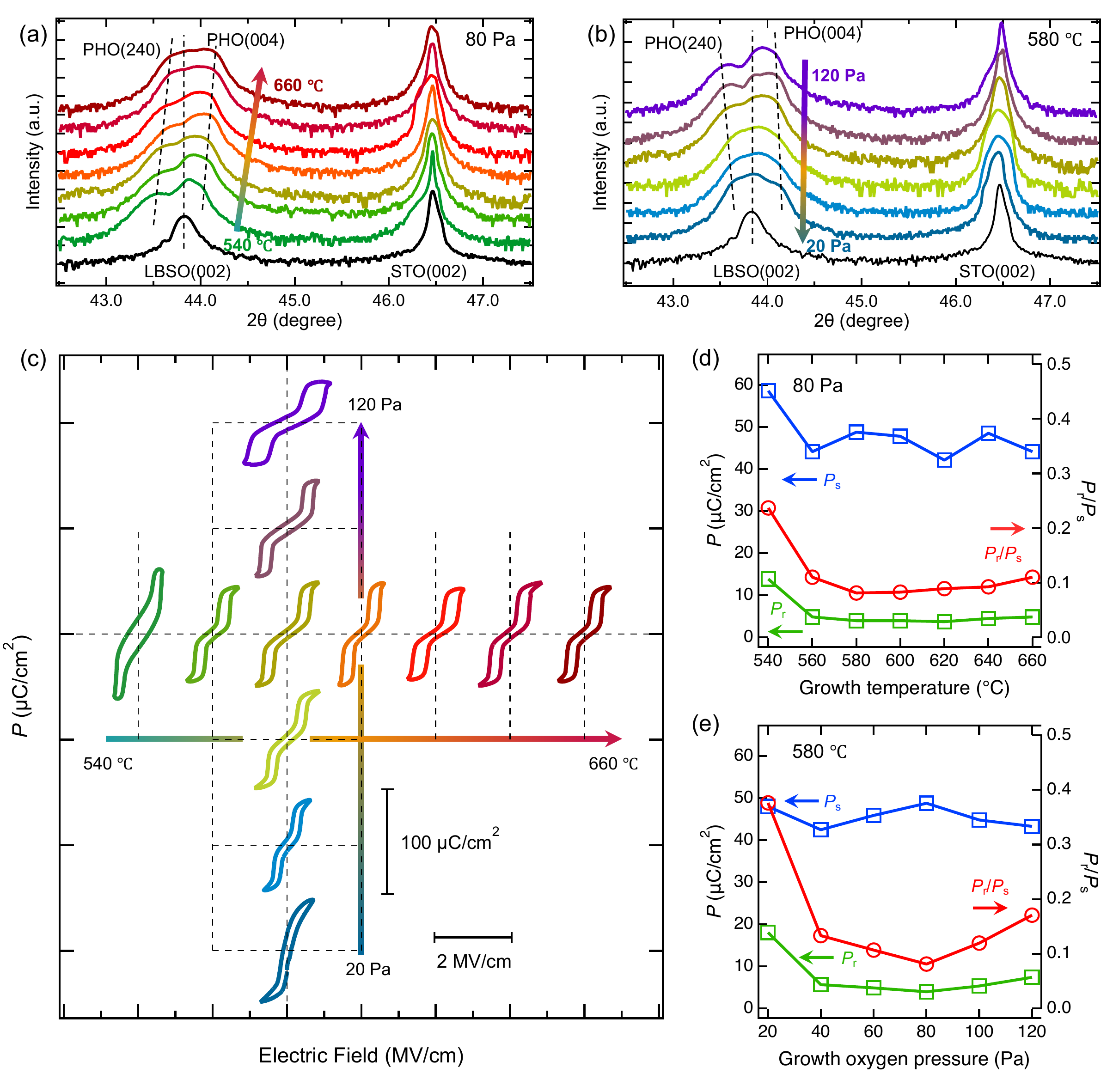}
\caption{(a,b) $2\theta$-$\omega$ scans of PHO/LBSO/STO(001) heterostructures synthesized with the substrate temperature ranging from 540 to 660 $^{\circ}$C and under an oxygen pressure between 20 and 120 Pa, respectively.  show a progressive movement of the PHO peak to lower and higher $2\theta$ angles with elevated growth pressure and temperature, respectively. Additionally, PHO(004) is more preferred at higher growth pressure and temperature conditions. The black line was obtained from a bare LBSO thin film, and a vertical offset was added to the XRD data for clarity. (c) the $P$-$E$ loops of the same samples in (a,b). Horizontal and vertical and offsets were added to the $P$-$E$ loops for clarity. Compared to PZO, the AFE of PHO demonstrates significantly greater resilience to variations in growth conditions, with the artificial FE becoming more pronounced at reduced growth pressures and temperatures. (d, e) $P_\text{r}$, $P_\text{s}$, and $P_\text{r}/P_\text{s}$ under different growth conditions.}
\label{PHO}
\end{figure*}

To confirm the generality of this phenomenon, we also fabricated thin films of another model AFE material, PHO, via PLD under different growth temperatures and oxygen pressures. PHO shares the similar crystal structure as PZO, with slightly smaller lattice parameters. To this end, the study of PHO thin films would serve as the most suitable control experiment. \ref{PHO}a and \ref{PHO}b show the $2\theta$--$\omega$ scan for PHO/LBSO/STO(001)$_\text{c}$ heterostructures (a broader spectrum of $2\theta$--$\omega$ scans is available in Figure S4, Supplementary Materials). 
XRD scans show PHO and LBSO peaks that are in close proximity to each other and that the epitaxial PHO thin films were coherently grown on the LBSO pseudo-substrates. 
{\color{black} Similar to the behavior of PZO thin films and earlier studies \cite{Weymann2020, Sarott2023}, the (240)$_\text{o}$ and (004)$_\text{o}$ shift to higher angles with increasing growth temperature, indicating decreased out-of-plane lattice parameter and the lattice volume (\ref{PHO}a). On the other hand, we find a slight lattice expansion with increased oxygen pressure (\ref{PHO}b), which seems to be contrary to the bombardment effect {\cite{Sahar2016}}, and could be linked to a competition of the Pb concentration gradient with the bombardment effect. Further efforts are needed to better understand this evolution.}

\begin{figure*}[ht]
\includegraphics[width=\linewidth]{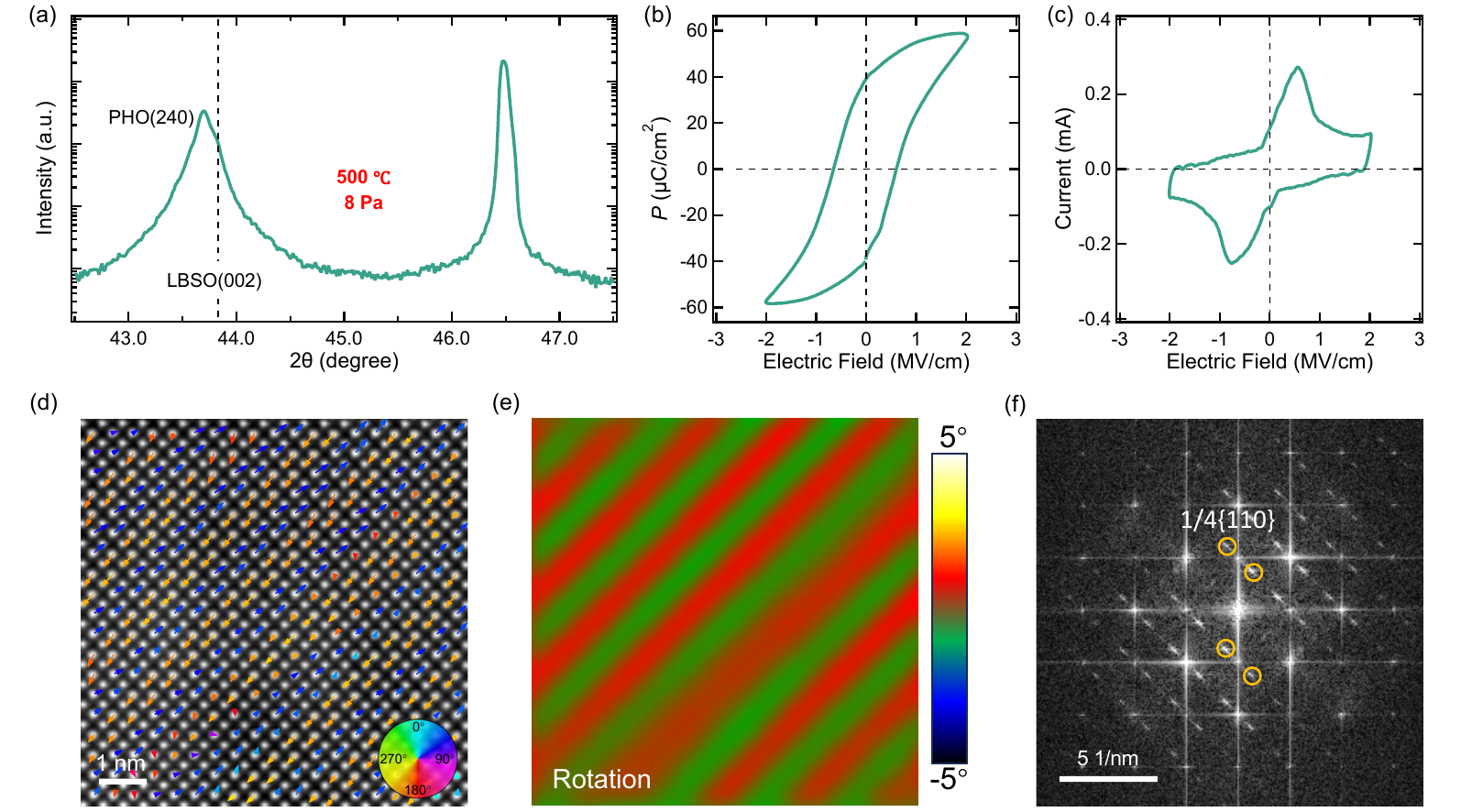}
\caption{(a) $2\theta$--$\omega$ scan for PHO/LBSO/STO(001) grown at 500 $^{\circ}$C and 8 Pa, exhibiting a dominated PHO(240) orientation. The $P$-$E$ loop (b) and the corresponding switching current curve (c), with an artificial FE feature. (d) The $\delta_\text{Pb}$ map in relation to its four closest Hf, overlaid on the corresponding HAADF--STEM image, exhibiting an AFE $\uparrow\uparrow\downarrow\downarrow$ dipole pattern. (e) The GPA lattice rotation map of (d) showing the same locally periodic structure. (f) FFT pattern featuring the satellite spots (circled in yellow) resulting from the same dipole arrangement.}
\label{FEPHO}
\end{figure*}

The $P$-$E$ loops associated with these heterostructures were then measured and depicted in \ref{PHO}c, while their $P_\text{r}$, $P_\text{s}$, and $P_\text{r}/P_\text{s}$ are summarized in \ref{PHO}d and \ref{PHO}e, respectively. The horizontal series in \ref{PHO}c illustrate the temperature-dependent $P$-$E$ of the PHO films at a constant growth oxygen pressure of 80 Pa.  Across the temperature range of 540 $^{\circ}$C to 660 $^{\circ}$C, well-defined double loops were achieved (the one at 540 $^{\circ}$C shows a leakage-caused broadening of $P$-$E$ loop), with the optimal one obtained at the growth temperature of 580 $^{\circ}$C. This is not exactly the same trend as observed in the PZO films but agrees with previous reports {\cite{Weymann2020, Sarott2023}} that the AFE nature and the double-hysteresis feature of PHO are more robust than those of PZO. Instead, we found that the oxygen pressure provides a more effective route to introduce strong disorder into the film and trigger a similar FE-like switching behavior in PHO films grown at a low oxygen pressure of 20 Pa (vertical column of \ref{PZO}c).  This evolution with growth pressure is linked to the structural disorder (such as point defects or defect complexes) that interact with the AFE-FE phase boundary and delay the phase transitions, as demonstrated in a similar growth pressure-dependent study of PZO films {\cite{Pan2023,Sahar2016}}.

\begin{figure*}[ht]
\includegraphics[width=\linewidth]{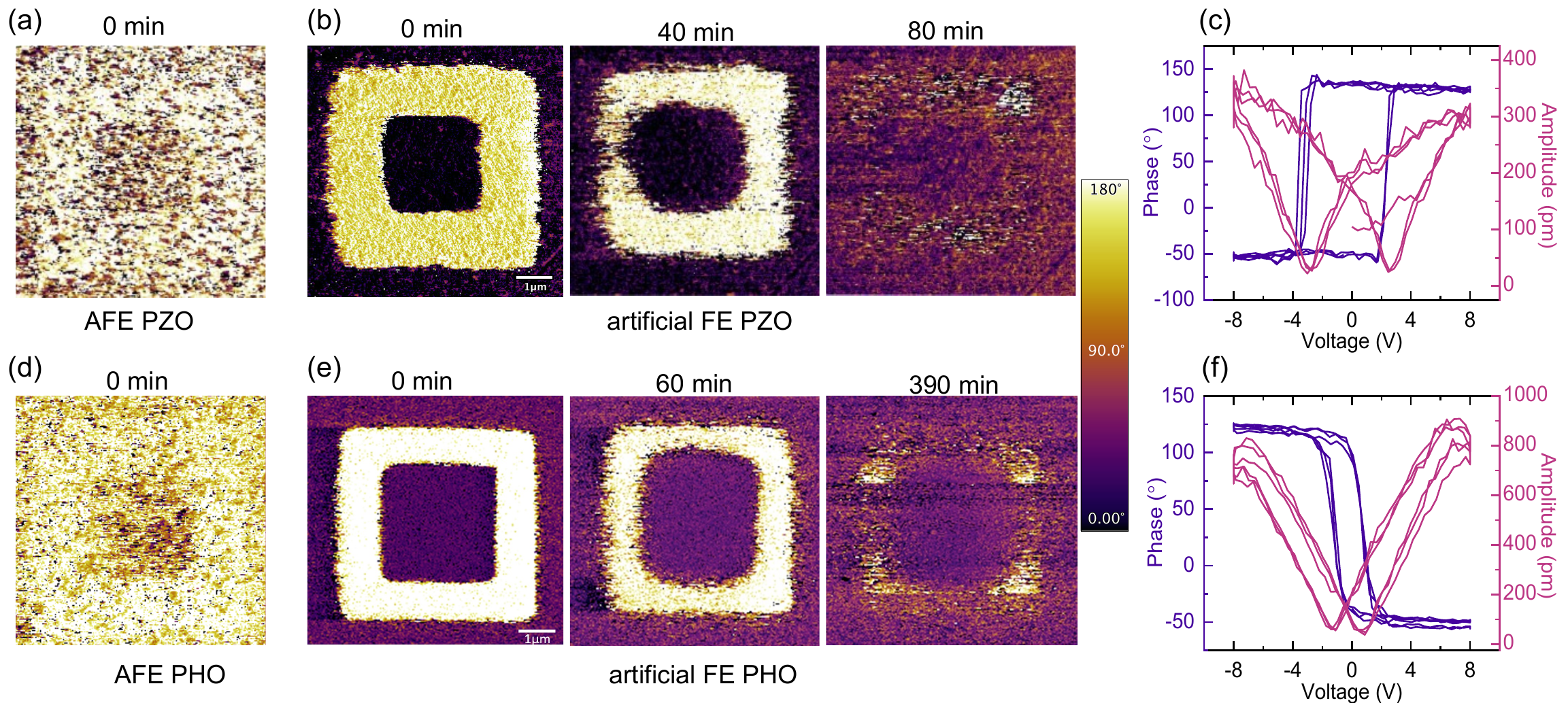}
\caption{Phase-contrast PFM images of AFE PZO (a) and PHO (d) after poling demonstrating weak phase contrast. Phase-contrast PFM images of FE PZO (b) and PHO (e) recorded at various time points after undergoing the same poling process. The poling state is maintained for hours before eventually returning to the AFE ground state. (c) and (f) are the phase and amplitude switching spectroscopy loops for FE-like PZO and PHO films.}
\label{PFM}
\end{figure*}

With this understanding, we fabricated a PHO thin film with an even lower pressure of 8 Pa and at a temperature of 500 $^{\circ}$C, which are conditions that are significantly suboptimal for growth, to induce greater disorder (i.e., a larger $t$ in \ref{Fig1}). The $2\theta$--$\omega$ XRD scan shows a dominating PHO(240) orientation (\ref{FEPHO}a), and as anticipated, the $P$-$E$ loop now exhibits an even more pronounced FE-like characteristic, with a remarkable hysteresis (\ref{FEPHO}b) and only two peaks in the polarization switching curves (\ref{FEPHO}c). The coercive field $E_\text{c}$ reaches 0.63 MV/cm for this PHO film grown at 8 Pa, much larger than the $E_\text{c}$ value (0.12 MV/cm) for the PHO film grown at 20 Pa. This enhancement can be ascribed to the increased intensity of disorder (complexity and concentration of defects) introduced into the film with lower growth pressure, which more strongly pins and delays the AFE-FE phase transitions {\cite{Pan2023}}. To support this speculation and to verify the AFE nature in this PHO film (instead of a seemingly FE phase as the $P$-$E$ loop shows), we employed the HAADF--STEM to analyze the local dipole configuration.  The map of Pb displacement ($\delta_\text{Pb}$) with respect to their four nearest Hf superimposed on the corresponding HAADF--STEM image was displayed in \ref{FEPHO}d. The $\uparrow\uparrow\downarrow\downarrow$ dipole pattern was distinctly observed, which is the typical structural feature observed in the AFE PZO and PHO.  {\cite{Liu2023,Dufour2023}} The geometric phase analysis (GPA) lattice rotation map of \ref{FEPHO}d was also derived from the HAADF--STEM image, which shows the typical locally periodic structure of orthorhombic AFE lattices.  From the STEM image, a fast Fourier transformation (FFT) pattern was further derived (\ref{FEPHO}f), where 1/4\{110\}$_\text{pc}$ satellite spots (marked by the yellow circles) are identified between the main lattice diffraction spots, corresponding to the quadrupling of the periodicity due to the Pb cation displacement (with the $\uparrow\uparrow\downarrow\downarrow$ pattern along \{110\}$_\text{pc}$, \ref{FEPHO}d). Therefore, despite the artificial FE-like single-hysteresis loops observed in the thin films of PHO (\ref{FEPHO}) and PZO (\ref{PZO}b), their ground state is found to remain AFE.{\color{black}Our examination of the structure through XRD and STEM analyses of the two AFE materials reveals no evidence of FE phases. This offers experimental validation for the hysteresis that mimics FE behavior in AFE materials, without invoking the presence of FE \cite{Chaudhuri2011, Gao2017} or ferrielectric \cite{Yao2023, Yu2024} phases.}{\color{black}Yet, it is important to observe that the local dipole configuration can deviate from the exact $\uparrow\uparrow\downarrow\downarrow$ pattern (\ref{FEPHO}d), indicating the existence of nonuniform disorder. Notably, the varying magnitudes of these Pb displacements and the antiphase boundaries also suggest nonuniform distributions of defects, local strain, and charge distribution, serving as extra origins of local disorder, which vary the pinning strength, smear the standard AFE double-hysteresis loop, and result in FE-like single loops.}

Next, we utilized the time-dependent PFM study to understand the stability and switching kinetics of the AFE and the artificial FE-like states in the PZO and PHO films. The detailed poling procedure is: a $5 \times 5$ $\upmu$m square region was poled-up by $+10$ V, and another $3 \times 3$ $\upmu$m square inside this region was further switched downward by a $-10$ V bias. For the prototypical AFE PZO and PHO films (\ref{PFM}a,d), their PFM phase contrast is nearly indistinguishable even right after poling with the high DC voltage of 10 V. This is consistent with the observations for $P$-$E$ loops (\ref{PZO}c and \ref{PHO}c), highlighting that in typical AFEs, field-induced FE polarization would not be retained after removal of the electric field, and spontaneously transforms back to the ground AFE state at a speed the PFM scan cannot catch (at the level of microsecond {\cite{Pan2024}}). In contrast, the FE-like PZO and PHO films exhibit substantially different behaviors. For example, the most remarkable FE-like thin films (PZO grown at 600 $^{\circ}$C; \ref{PZO}c, and PHO grown at 8 Pa; \ref{FEPHO}b) were poled by the same poling procedure. As shown in \ref{PFM}b,e, there is a 180$^{\circ}$ phase reverse between the two square areas right after applying the bias voltages, with the dark and bright areas indicating downward and upward polarization, respectively, exhibiting an FE-like behavior. However, in real FE films, it should be noted that the FE ground state is nonvolatile and would have a theoretically infinite relaxation time. Yet it was found that in these FE-like PZO and PHO films, the poled phase contrast gradually reduces and almost fully disappears after a relaxation period ($\sim$80 min for the PZO film and $\sim$390 min for the PHO film). This clearly suggests that the field-induced FE polarization in these films is not at the ground state; in other words, the polarization and the FE phase are metastable and are just temporarily pinned by the growth-introduced structural disorder in the materials, which would eventually transform back to the AFE ground state at room temperature.

\section{Conclusion}
To summarize, the disorder level in the AFE PZO and PHO films was systematically manipulated by varying the growth conditions, based on the LBSO pseudo-substrate that allows for dislocation-free epitaxial growth. A progressive enhancement in the disorder strength, accompanied by an expansion of its distribution (i.e., non-uniform disorder), can be achieved by deviating from the optimal growth conditions. Consequently, the disordered AFE $P$-$E$ loops superimpose to smear the typical double-hysteresis into artificial FE-like single-hysteresis $P$-$E$ loops. The stability of the metastable FE state over a duration of hours is noted. Our research highlights the innovative influence of disorder in altering the polarization characteristics in AFEs, which provides a novel pathway for new or enhanced functionalities.

% \medskip
% \textbf{Declaration of competing interest} \par %delete if not applicable))
% The authors declare that they have no known competing financial interests or personal relationships that could have appeared to
% influence the work reported in this paper.

% Acknowledgements

\medskip
\noindent
\textbf{Acknowledgements} \par %delete if not applicable))
Y.Z., X.Z. and Z.Z contributed equally to this work. L.W. acknowledges funding through the Xingdian Talent Support Project of Yunnan Province (Grant No. KKRD202251009) and Yunnan Fundamental Research Projects (Grant Nos. 202101BE070001-012 and 202201AT070171). J.G. acknowledges the support from the Natural Science Foundation of China (Grant No. 52001117).

 \bibliographystyle{elsarticle-num} 
 \bibliography{ref}





% \end{thebibliography}
\end{document}